\documentclass[aps,prd,nofootinbib,twocolumn,amsmath,amssymb,longbibliography]{revtex4-1}

\usepackage{amsmath,amssymb,amsfonts,latexsym,cancel}

\usepackage{mathtools}
\usepackage{graphicx}
\usepackage{color}
\usepackage{amsbsy}
\usepackage{hyperref}
\usepackage{tikz}

\usepackage{seqsplit}

\usepackage{comment}

\newcommand{\be}{\begin{equation}}
\newcommand{\beq}{\begin{equation}}
\newcommand{\ee}{\end{equation}}

\newcommand{\ba}[1]{\begin{align}#1\end{align}}

\def\bea {\begin{eqnarray}}
\def\eea {\end{eqnarray}}

\def\f{\frac}

\def\nn{\nonumber}
\def\lp{\ell_{\rm Pl}}

\def\dd{{\rm d}}

\def\mH{\mathcal{H}}

\begin{document}

\title{Static Planck stars from effective loop quantum gravity}

\author{Edward Wilson-Ewing} \email{edward.wilson-ewing@unb.ca}
\affiliation{Department of Mathematics and Statistics, University of New Brunswick, \\
Fredericton, NB, Canada E3B 5A3}

\begin{abstract}

Effective loop quantum gravity dynamics are derived for spherically symmetric spacetimes with a perfect fluid matter content. For homogeneous spacetimes, the effective dynamics agree with the standard results of loop quantum cosmology, while the equations for static solutions give an effective Tolman-Oppenheimer-Volkoff equation. There exist solutions to the effective Tolman-Oppenheimer-Volkoff equation that have a mass of the order of the Planck mass, a Planckian radius, and no horizon; these miniature stars could potentially contribute to dark matter, and could be an end state for an evaporating black hole.

\end{abstract}

\maketitle

\section{Introduction}

Quantum gravity is widely believed to have an important role to play in the foundations of astrophysics.

First, quantum gravity may resolve the singularities that arise in classical general relativity, whether inside black holes or in the early universe \cite{Penrose:1964wq, Hawking:1970zqf}. Although the interior of a black hole cannot be probed by observers who remain outside its horizon, it is clearly desirable to have a complete description of black holes that is free of pathologies like singularities, as is generally expected to arise from quantum gravity; it is also thought that quantum gravity will resolve the black hole information problem \cite{Hawking:1975vcx}. Similarly, a quantum description of the universe may resolve the big-bang singularity, and potentially provide a quantum bridge to a pre-big-bang universe as predicted, for example, by loop quantum cosmology \cite{Ashtekar:2006rx}.

Second, quantum gravity may give some important insights into the source of dark matter or of dark energy. For example, are quantum black hole remnants components of dark matter? And does the quantum vacuum energy contribute to dark energy?

Effective frameworks can be very useful in investigating some of these questions. For loop quantum gravity \cite{Ashtekar:2004eh, Rovelli:2004tv, Thiemann:2007pyv}, effective dynamics typically include effects due to the Planck-scale discrete quantum geometry, but neglect quantum fluctuations \cite{Bojowald:2005cw, Ashtekar:2006wn, Taveras:2008ke}; this is expected to be a good approximation when studying physics at length scales large compared to the Planck length (even when the spacetime curvature is Planckian) \cite{Rovelli:2013zaa, Bojowald:2015fla}. In cases that quantum fluctuations become large, it is of course necessary to go beyond these effective frameworks, but even in such cases effective models can be expected to provide a useful first-order approximation \cite{Bojowald:2008ec, Diener:2014mia}.

The focus of this paper is to consider spherically symmetric spacetimes with a perfect fluid matter content. These spacetimes are of considerable interest as, on the one hand, they are sufficiently simple that calculations are tractable and, on the other hand, are quite rich as they allow for local degrees of freedom and can describe both cosmological and black hole spacetimes, including collapse models where a black hole forms dynamically. As a result, they can give insight into the possibility of quantum gravity curing black hole and cosmological singularities, as well as provide a testing ground for dark matter models based on quantum black holes or black hole remnants.

There has already been some work studying spherically symmetric spacetimes coupled to matter with a local degree of freedom in effective loop quantum gravity, but these have all focused on simple matter fields, either dust (i.e., the Lema\^itre-Tolman-Bondi spacetimes) \cite{Bojowald:2008ja, Bojowald:2009ih, Tibrewala:2012xb, Liu:2014kra, Kelly:2020lec, Alonso-Bardaji:2021tvy, Husain:2022gwp, Giesel:2023tsj, Alonso-Bardaji:2023qgu}, or a scalar field \cite{Bojowald:2015zha, BenAchour:2016brs, Benitez:2020szx, Alonso-Bardaji:2020rxb, Gambini:2021uzf}. (There has also been considerable work studying vacuum solutions, see, e.g., \cite{Perez:2017cmj, Gambini:2022hxr, Ashtekar:2023cod} for recent reviews of this large body of work.) The main goal of this paper is to generalize the earlier work for any perfect fluid; in particular, this generalization is important in finding static star-like solutions, where the pressure must grow inside the star to counteract the gravitational attraction.

As shall be shown in more detail below, the generalization to allow for any perfect fluid matter content can be obtained by adding a term to the Hamiltonian constraint, and many of the mathematical tools already developed in previous work will be of use here. Nonetheless, the physics is considerably richer and allows for interesting new solutions.

The generalized Painlev\'e-Gullstrand coordinates are a convenient choice of coordinates for a Hamiltonian analysis in spherical symmetry. These coordinates are based on the areal gauge \cite{Kuchar:1994zk} where the radial coordinate is chosen to be (the square root of) the prefactor to $\dd\Omega^2$ in the metric, namely
\beq \label{pg-coords}
\dd s^2 = -N^2 \dd t^2 + \f{1}{1+\epsilon} \Big( \dd r + N^r \dd t \Big)^2 + r^2 \dd \Omega^2.
\ee
The remaining components of the metric are the lapse $N$, the radial component of the shift vector $N^r$ (the other components of the shift vector vanish due to spherical symmetry), and $\epsilon > -1$, which all depend on $t$ and $r$.  A useful shorthand is $\dot f = \partial f / \partial t, ~ f' = \partial f / \partial r$.

Assuming the matter content is given by a perfect fluid $T_{ab} = (\rho + p) u_a u_b + p g_{ab}$, the conservation of the stress-energy tensor projected on the plane orthogonal to $u^a$, namely $(\delta^b_c + u^b u_c) \nabla^a T_{ab} = 0$, gives
\beq \label{lapse-pressure}
\f{N'}{N} = - \f{p'}{\rho+p}.
\ee
This form of \eqref{lapse-pressure} assumes $p \neq -\rho$; if $p = -\rho$ then $p' = 0$. Note that this relation only relies on the form of the metric \eqref{pg-coords}, the conservation of the stress-energy tensor, and on the fact that the matter content is a perfect fluid---importantly, it is independent of any equations of motion and will hold in any effective framework where $\nabla^a T_{ab} = 0$, as is the case for effective LQG models.

\section{Effective Dynamics}

In terms of Ashtekar-Barbero variables, in the areal gauge the metric has the general form \cite{Kelly:2020uwj}
\beq
\dd s^2 = -N^2 \dd t^2 + \f{(E^b)^2}{r^2} ( \dd r + N^r \dd t )^2 + r^2 \dd\Omega^2.
\ee
Here $E^b$ is the component of the densitized triad pointing in angular directions, and the shift vector is
\beq \label{def-shift}
N^r = -\f{N r}{\gamma \sqrt\Delta} \sin \f{\sqrt\Delta b}{r} \cos \f{\sqrt\Delta b}{r},
\ee
where $b$ is the angular component of the Ashtekar-Barbero connection that is canonically conjugate to $E^b$,
\beq
\{b(r_1), E^b(r_2)\} = G \gamma \delta(r_1 - r_2),
\ee
$\gamma$ is the Barbero-Immirzi parameter and $\Delta \sim \lp^2$ is the area gap, the smallest non-zero eigenvalue of the LQG area operator. Clearly, $\epsilon$ in the metric \eqref{pg-coords} and $E^b$ are related by $E^b = r/\sqrt{1+\epsilon}$. The form of the shift \eqref{def-shift} is required by gauge-consistency conditions \cite{Giesel:2023hys}; this can also be understood by considering the cosmological sector since in this case the shift $N^r$ (up to a prefactor of $-Nr$) is given by the Hubble rate, that in effective loop quantum cosmology is $(\gamma \sqrt\Delta)^{-1} \sin \f{\sqrt\Delta b}{r} \cos \f{\sqrt\Delta b}{r}$ \cite{Husain:2022gwp}.

With this gauge-fixing, in spherical symmetry the effective dynamics that include holonomy corrections as motivated by loop quantum gravity can be derived from the effective scalar constraint:
\ba{
\mH =& \, \Bigg[ - \f{E^b}{2G\gamma^2\Delta r} \partial_r \left( r^3 \sin^2 \f{\sqrt\Delta b}{r} \right) - \f{1}{2G} \left( \f{E^b}{r} - \f{r}{E^b} \right) \nn \\ & \qquad
+ \f{r}{G} \partial_r \left( \f{r}{E^b} \right) + \mH_m \Bigg].
}
This effective constraint was initially considered for the vacuum case \cite{Kelly:2020uwj}, the only difference here is the addition of the Hamiltonian density for the perfect fluid  $\mH_m$.  Note that the diffeomorphism constraint has been solved by imposing the areal gauge (before including LQG effects), so the only remaining constraint is the Hamiltonian constraint $\mH$. (Differently from the case of dust, the lapse $N$ is left free as the dust-time gauge is not used here.)

The energy density and pressure of the perfect fluid can be calculated from $\mH_m$ through \cite{Singh:2009mz}
\beq
\rho = \f{\mH_m}{V} = \f{\mH_m}{4 \pi r E^b}, \quad p = - \, \f{\delta \mH_m}{\delta V} = - \, \f{1}{4 \pi r} \cdot \f{\delta \mH_m}{\delta E^b},
\ee
where $V$ is the spatial volume element (and $\delta \mH_m / \delta b = 0$).

From $\mH = 0$, it then follows that
\beq
\rho = \f{1}{8 \pi Gr^2} \, \partial_r \left( \f{r^3}{\gamma^2\Delta} \sin^2 \f{\sqrt\Delta b}{r} - r \epsilon \right),
\ee
which suggests defining the gravitational mass $m(r)$ as
\beq \label{def-m}
m(r) = 4 \pi \int_0^r \!\! \dd \tilde r~ \tilde r^2 \rho = \f{1}{2G} \left( \f{r^3}{\gamma^2\Delta} \sin^2 \f{\sqrt\Delta b}{r} - r \epsilon \right).
\ee
Note that the gravitational mass is different from the proper mass $m_p = 4 \pi \int_0^r \! \dd \tilde r \,\, (1+\epsilon)^{-1/2} \tilde r^2 \rho(\tilde r, t)$ whose integral includes the contribution to the spatial determinant from the $g_{rr}$ component of the metric.

The equations of motion then follow from $\dot f = \{f, H\}$, where $H = \int \dd r \, N \mH$ is the Hamiltonian,
\ba{
\label{bdot}
\dot b =& \, - \f{N}{2 \gamma \Delta r} \partial_r \left( r^3 \sin^2 \f{\sqrt\Delta b}{r} \right) + \f{\gamma N \epsilon}{2r} \nn \\ & \qquad
- \f{\gamma N (1 + \epsilon) p'}{\rho + p} - 4 \pi G \gamma N r p, \\
\label{edot}
\dot E^b =& \, N \Bigg[ -\f{1}{\gamma\sqrt\Delta} r^2 \partial_r \left( \f{E^b}{r} \right) \sin \f{\sqrt\Delta b}{r} \cos \f{\sqrt\Delta b}{r} \nn \\ & \qquad
+ \f{1}{\gamma \sqrt\Delta} r E^b \sin \f{\sqrt\Delta b}{r} \cos \f{\sqrt\Delta b}{r} \f{p'}{\rho+p} \Bigg],
}
using $N' = -Np'/(\rho+p)$ from \eqref{lapse-pressure}.

It is convenient to rewrite the equation of motion for $E^b$ in terms of $\epsilon$,
\beq
\dot \epsilon = - \f{N r}{\gamma \sqrt\Delta} \sin \f{\sqrt\Delta b}{r} \cos \f{\sqrt\Delta b}{r} \left( \epsilon' + \f{2(1+\epsilon)}{\rho+p} p' \right),
\ee
and it is also illuminating to calculate the equation of motion for $m$ from the definition \eqref{def-m},
\beq
\dot m = - \f{N r}{\gamma \sqrt\Delta} \sin \f{\sqrt\Delta b}{r} \cos \f{\sqrt\Delta b}{r} \Big( m' + 4 \pi r^2 p \Big),
\ee
note that the prefactor for both of these equations is $N^r$. The correct classical equations of motion are obtained in the limit $\Delta \to 0$, cf.\ \cite{Lasky:2006mg}.

The basic equations of motion are those for $b, E^b$ (or equivalently $\epsilon$ instead of $E^b$), combined with a method to determine the pressure (typically through an equation of state relating $p$ to $\rho$) and the relation \eqref{lapse-pressure} to determine the lapse $N$ at each instant of time.  From these, it is possible to calculate $m$, $N^r$, and all other geometric and matter quantities for the spacetime.

Note that these effective dynamics are derived within a gauge-fixed formalism due to the use of the generalized Painlev\'e-Gullstrand coordinates. Although this set of coordinates is convenient as it significantly simplifies the problem, a drawback of gauge-fixing is that it is not possible to directly check the covariance of the effective dynamics (for recent work on this question in vacuum spherical symmetry, see \cite{Alonso-Bardaji:2023vtl, Zhang:2024khj, Belfaqih:2024vfk}). This check will require a more general analysis that avoids selecting any particular set of coordinates and is left for future work.

\section{Cosmology}

The homogeneous and isotropic FLRW spacetimes are recovered for the case that $\rho$ and $p$ are independent of $r$. Then, $N$ is also independent of $r$ and can be any function of time; for convenience it is set to $N=1$ here.

The FLRW metric in generalized Painlev\'e-Gullstrand coordinates (with $N=1$) is
\beq
\dd s^2 = -\dd t^2 + \f{1}{1-k r^2 / a^2} \Big( \dd r - rH \dd t \Big)^2 + r^2 \dd \Omega^2,
\ee
where $a$ is the scale factor, $H = \dot a / a$ is the Hubble rate and $k=$ constant is the spatial curvature.  It follows directly that $\epsilon = -kr^2/a^2$ and $N^r = -rH$.  (This form of the FLRW metric can be obtained from the more common diagonal form $\dd s^2 = -\dd t^2 + a^2 [ (1-k \chi^2)^{-1} \dd \chi^2 + \chi^2 \dd \Omega^2]$ through the coordinate transformation $r = a \chi$.)

This identification shows that
\beq
H = \f{1}{\gamma \sqrt\Delta}  \sin \left(\f{\sqrt\Delta b}{r}\right) \cos \left(\f{\sqrt\Delta b}{r}\right),
\ee
and the equation for $\dot m$ becomes the continuity equation
\beq
\dot \rho = -3H (\rho + p),
\ee
while the equation for $\dot \epsilon$ is automatically satisfied.

The effective Friedman equation comes from \eqref{def-m},
\ba{
H^2 & \, = \f{1}{\gamma^2 \Delta}  \sin^2 \left(\f{\sqrt\Delta b}{r}\right) \cos^2 \left(\f{\sqrt\Delta b}{r}\right) \nn \\
& \, = \left( \f{8 \pi G}{3} \rho - \f{k}{a^2} \right) \cdot \left( 1 - \f{\rho}{\rho_c} + \f{\gamma^2 \Delta k}{a^2} \right),
}
which exactly agrees with the effective dynamics for the so-called K quantization of loop quantum cosmology (LQC) \cite{Vandersloot:2006ws, Singh:2013ava}, for any spatial curvature and any pressure; here $\rho_c = 3 / 8 \pi G \gamma^2 \Delta$ is the LQC critical energy density.

The solution for de Sitter space is obtained for $p = -\rho$. Then, \eqref{lapse-pressure} has to be rewritten as $p'=0$, and this in turn implies $\rho'=0$: in spherical symmetry the equation of state $p = -\rho$ imposes homogeneity.  In terms of Painlev\'e-Gullstrand coordinates, in classical general relativity $\epsilon = 0$ in de Sitter space, so it is sufficient to look for solutions with $\epsilon = 0$ in this context as well (note that this condition, together with the equation of state $p = -\rho$ ensures that $\dot m = \dot \epsilon = 0$), in which case
\beq
b = - \f{r}{\sqrt\Delta} \arcsin \sqrt{ \f{8 \pi G \gamma^2 \Delta}{3} \rho}.
\ee
Due to homogeneity it is possible to take $N=1$, then
\ba{
N^r = & \, - \f{r}{\gamma \sqrt\Delta} \sin \f{\sqrt\Delta b}{r} \cos \f{\sqrt\Delta b}{r} \nn \\ &
= r \, \sqrt{ \f{8 \pi G}{3} \rho \left(1 - \f{\rho}{\rho_c} \right) }.
}
In terms of the cosmological constant $\Lambda = 8 \pi G \rho$, the metric is (see also \cite{Lin:2024flv})
\beq
\dd s^2 = - \dd t^2 + \left( \dd r + r \sqrt \f{\Lambda}{3} \sqrt{1 - \f{\gamma^2 \Delta \Lambda}{3}} \dd t \right)^2 + r^2 \dd \Omega^2.
\ee
This solution approaches the de Sitter solution of general relativity in the limit of $\Delta \to 0$, and is very close to it for small $\Lambda$, as is the case for our universe.

As an aside, it is worth pointing out that in the case that $\Lambda$ is Planckian and only slightly smaller than $3 / \gamma^2 \Delta$, then it is also possible to have a slow expansion rate, as is observed in our universe. Can this solve the cosmological constant problem by combining a Planck-scale $\Lambda$ with the observed rate of expansion? The difficulty with this solution is that, since in such a scenario $\Lambda > 3 / 2 \gamma^2 \Delta$, it follows that at earlier times with a larger energy density (and closer to the bounce at $\rho_c$ where $H=0$), the Hubble rate will be smaller, not larger, and the Hubble rate would increase with time in an expanding universe.

\section{Static Solutions}

Static solutions are also of considerable interest as they may approximate long-lived astrophysical objects. Such configurations have a vanishing shift vector $N^r=0$, which from \eqref{def-shift} and \eqref{def-m} gives
\beq \label{zero-shift}
\epsilon = -\f{2Gm}{r}, \qquad {\rm or} \qquad \epsilon = -\f{2Gm}{r} + \f{r^2}{\gamma^2\Delta},
\ee
and also that $\dot m = \dot\epsilon = 0$. The first possibility in \eqref{zero-shift} is exactly the classical relation, so here I will focus on the second possibility which arises solely due to LQG effects.

Requiring that $\dot N^r = 0$ so the condition $N^r = 0$ is dynamically preserved, using \eqref{def-m} and \eqref{bdot}, implies that
\ba{ \label{dot-shift}
\dot N^r = & \, \f{N^2}{2} \left( \cos^2 \f{\sqrt\Delta b}{r} - \sin^2 \f{\sqrt\Delta b}{r} \right) \nn \\ & \qquad
\cdot \left( \f{2Gm'}{r} + 8 \pi G r p + \epsilon' + \f{2(1+\epsilon)}{\rho+p} p' \right),
}
must vanish (note that the $\dot N$ term does not contribute as it is multiplied by a function that vanishes when $N^r=0$), giving
\beq \label{tov}
\f{2Gm'}{r} + 8 \pi G r p + \epsilon' + \f{2(1+\epsilon)}{\rho+p} p' = 0,
\ee
since exactly one of $\sin (\sqrt\Delta b /r)$ and $\cos (\sqrt\Delta b / r)$ vanishes when $N^r=0$.
This is the Tolman-Oppenheimer-Volkoff equation for the pressure in static spherically symmetric configurations; interestingly, it has exactly the same form for the LQG effective dynamics as in classical general relativity. There is a key difference though---for the quantum solutions of interest here, it is the second equality in \eqref{zero-shift} that is used rather than the first one (relevant for classical general relativity).  As a result, substituting  $\epsilon = - 2Gm/r + r^2/(\gamma^2\Delta)$ in \eqref{tov} gives
\beq \label{lqg-tov}
p' = - \f{\rho + p}{1 - \f{2Gm}{r} + \f{r^2}{\gamma^2\Delta}} \left( \f{Gm}{r^2} + 4\pi G r p + \f{r}{\gamma^2\Delta} \right).
\ee
Combining this with an equation of state $p = f(\rho)$, or some other way to fix one of $\rho$ and $p$, as well as the radial equation for the lapse $N$ \eqref{lapse-pressure} (that is the same form as in classical general relativity), completes the set of equations to solve for a static configuration. Note that if there is a boundary at some fixed $r=R$ with a vacuum region outside, $p(R)=0$ is required for continuity, just as in general relativity.

\subsection{Constant Energy Density of $\rho_c$}

An especially simple configuration is the case of a star of constant density,
\beq \label{const-dens}
\rho = \begin{cases} \rho_c \qquad &{\rm for} ~ r < R, \\ 0 \qquad &{\rm for} ~ r > R. \end{cases}
\ee
where $\rho_c$ is critical energy density
\beq
\rho_c = \f{3}{8 \pi G \gamma^2 \Delta}.
\ee
Although this example is somewhat artificial (it would be more realistic to specify boundary conditions and an equation of state, and then integrate the LQG effective Tolman-Oppenheimer-Snyder equation inwards from the outer boundary of the star), it is of interest due to its simplicity which allows for an analytic solution.

For this configuration, the LQG-corrected Tolman-Oppenheimer-Volkoff equation simplifies considerably to
\beq
p' = -4\pi G (p + \rho_c)^2 r,
\ee
which can be integrated by separation of variables to give
\beq
p = \f{1}{2\pi G r^2 + C} - \rho_c,
\ee
where $C$ is the constant of integration that is fixed by requiring that the pressure at the surface of the star vanishes, $p(R) = 0$.  This gives
\beq
p = \f{2\pi G \rho_c^2 (R^2-r^2)}{1 - 2\pi G \rho_c (R^2-r^2)},
\ee
note that solutions are only possible for the case that $R < 1/\sqrt{2 \pi G \rho_c} = 2 \gamma \sqrt{\Delta/3}$; this condition implies that stars of constant density $\rho_c$ must satisfy
\beq \label{max-mass}
M < \f{4 \gamma \sqrt\Delta}{\sqrt{27} G}.
\ee
This mass is sufficiently small so no horizons form \cite{Kelly:2020uwj}.

The metric for the vacuum exterior (in Painlev\'e-Gullstrand coordinates with LQG corrections) is \cite{Kelly:2020uwj}
\beq
\dd s^2 = - \dd t^2 + \left( \dd r + \sqrt{ \f{r_S}{r} \left( 1 - \f{\gamma^2 \Delta r_S}{r^3} \right)} \dd t \right)^2 + r^2 \dd \Omega^2,
\ee
where $r_S = 2GM$ is the Schwarzschild radius, with $M = m(R) = 4 \pi R^3 \rho_c / 3$. It is easy to verify that at $r=R$, the coefficient to the $\dd r^2$ term matches for the interior and exterior solutions, being exactly 1 in both cases; and the coefficient to the cross term $\dd t \, \dd r$ vanishes at $r=R$ for the exterior solutions (and remains 0 for $r \le R$).

Finally, it is a straightforward calculation to solve for the lapse $N$ from \eqref{lapse-pressure}. Using the boundary condition that $N(R) = 1$ to ensure the metric is continuous, the result for $r \le R$ is
\beq
N = 1 - 2\pi G \rho_c (R^2-r^2).
\ee
As mentioned above, the configuration studied here is a simple one that is not particularly realistic, but it has the advantage that it can be solved analytically. More realistic solutions can be derived by numerically integrating the effective Tolman-Oppenheimer-Volkoff equation from the outer boundary for a given equation of state; although the details can be expected to differ, it seems likely that there will continue to exist static solutions with a total mass of the order of $m_{\rm Pl}$. Nonetheless, due to the simplicity of the particular solution derived here, the following discussion on potential physical consequences is tentative, and is based on the assumption of the existence of more realistic static solutions with a similar total mass.

These gravitationally static miniature stars (or `Planck stars', borrowing a term that was initially used for a different, bouncing solution \cite{Rovelli:2014cta}) of Planckian radius and density provide a potential candidate for dark matter: composite (horizonless) bodies of total mass $\sim m_{\rm Pl}$.

Bodies of mass $\sim m_{\rm Pl} \approx 10^{-5}$~g are interesting candidates for dark matter (or perhaps a portion of dark matter) \cite{Chung:1998zb, Kuzmin:1998uv}. Direct detection of such a dark matter candidate today seems challenging, although one possibility is that if two Planck stars coalesce so that the total mass is greater than \eqref{max-mass}, then the resulting body would presumably rapidly fall apart in an energetic process that could potentially be observed (this is similar to the idea of Planck-mass black hole remnants coalescing and then rapidly emitting high-temperature Hawking radiation \cite{Barrau:2019cuo}), while another possibility could be through gravitational quantum interference \cite{Christodoulou:2023hyt}. As an alternative to direct detection, constraints on Planck stars as dark matter are perhaps most likely to be obtained by considering their history.

For example, one possible way to create such a static Planck star could be as the end state of a black hole evaporating from Hawking radiation (at least in the case of zero angular momentum). Note that the maximal mass of this solution \eqref{max-mass} is half the mass at which the horizon disappears \cite{Kelly:2020uwj}, and as a result one may expect a highly dynamical process to occur when the horizon disappears and evaporation ends, for the mass to decrease by a factor of two to reach the static configuration; the study of this process is left for future work.

Therefore, Planck stars could potentially be directly formed in the very early universe, or arise as remnants of evaporated primordial black holes (see, e.g., \cite{MacGibbon:1987, Carr:2021bzv} for general discussions on primordial black holes and Planckian remnants, and \cite{Rovelli:2018hbk, Rovelli:2018hba, Rovelli:2018okm, Barrau:2021spy, Amadei:2021aqd, Papanikolaou:2023crz, Barca:2023shv, Borges:2023fub} for proposals within the context of LQG). In an inflationary scenario the first possibility would not significantly contribute to dark matter since the Planck stars would presumably be formed at near-Planckian energy scales, before inflation starts, and they would therefore be strongly diluted by the inflationary expansion; the second possibility is also constrained in an inflationary setting \cite{Carr:2021bzv}, although constraints may be weaker in other early-universe scenarios than inflation.

Also, although these Planck star solutions are static, it is not clear if they are thermodynamically stable. This will depend in particular on the temperature of such a body---if the temperature is Planckian, then these bodies will very rapidly evaporate through black-body radiation. On the other hand, if the temperature is sufficiently low, perhaps due to the gravitational field being so strong and the constituent particles being tightly bound with a very low kinetic energy, then these solutions will be thermodynamically stable. Further work is required to answer this question. Another question concerns the applicability of a fluid description to this particular solution. Under the assumptions that many particles form the Planck star and their mean free path is vanishingly small, a fluid approximation is viable even at these scales. It would nonetheless be desirable to go beyond this approximation in future work.

It is important to emphasize that there exist other static solutions to \eqref{lqg-tov}, including perturbations to the configuration \eqref{const-dens}, but it is typically difficult to derive exact analytical solutions; the study of these other solutions is left for future work. Note as well that there exist uniform density solutions satisfying the first condition of \eqref{zero-shift} that can also give Planck star solutions; the condition that the pressure remain finite in this case gives the usual Buchdahl bound $M \le 4R/9G$ \cite{Wald:1984}---for an interior density of $\rho_c$ the maximum mass possible is $M = 8 \gamma \sqrt{2 \Delta} / 9G$.

\section{Discussion}

In summary, the effective LQG equations for spherically symmetric gravity coupled to a perfect fluid can be derived from the effective Hamiltonian developed for the vacuum case \cite{Kelly:2020uwj} by adding a Hamiltonian density for a perfect fluid. In the cosmological sector, these equations of motion reduce to the loop quantum cosmology Friedman equations, and for static configurations it is possible to derive a Tolman-Oppenheimer-Volkoff equation for effective LQG that allows solutions for bodies of Planckian density and total mass $\sim m_{\rm Pl}$.

These static Planck stars could be a dark matter candidate if they are also thermodynamically stable. In addition, these solutions could also be the end state of an evaporating black hole: as they do not have a horizon, they do not emit any Hawking radiation.

A next step for future work would be to apply the effective equations of motion \eqref{bdot}--\eqref{edot} to other spherically symmetric systems beyond homogeneous or static solutions. In particular, it would be interesting to study configurations that collapse and dynamically form a black hole. Previous work studying black hole collapse with local degrees of freedom in effective LQG have so far only considered the cases where the collapsing matter is either dust \cite{Husain:2021ojz, Husain:2022gwp, Giesel:2023hys, Alonso-Bardaji:2023qgu, Fazzini:2023ova, Cafaro:2024vrw, Cipriani:2024nhx} or a massless scalar field \cite{Benitez:2020szx}, and it is important to consider other, more realistic, types of matter; a step in this direction could be to study other types of perfect fluids using these effective dynamics.

\medskip

\acknowledgments

\noindent
I thank Viqar Husain and Carlo Rovelli for helpful comments on an earlier draft of the paper.
This work was supported in part by the Natural Sciences and Engineering Research Council of Canada.

\raggedright

\end{document}